\documentclass[
reprint,
superscriptaddress,
showpacs,
amsmath,
amssymb,
aps,
pra,
longbibliography
]{revtex4-1}

\usepackage{graphicx}
\usepackage{color}
\usepackage{hyperref}
\usepackage{ulem}

\newif\ifcmnt
\cmnttrue 
\ifcmnt
    \providecommand{\aucmnt}[1]{#1}
    
\else
    \providecommand{\aucmnt}[1]{}
    
\fi

\newcommand{\rhotrue}{\rho_{\text{true}}}

\begin{document}


\title{Quadrature Histograms in Maximum Likelihood Quantum State Tomography}
\author{J. L. E. Silva}
\affiliation{Departamento de Engenharia de Teleinform\'atica, Universidade Federal do Cear\'a, Fortaleza, Cear\'a, 60440, Brazil}
\author{S. Glancy}
\affiliation{Applied and Computational Mathematics Division, National Institute of Standards and Technology, Boulder, Colorado, 80305, USA}
\author{H. M. Vasconcelos}
\email{hilma@ufc.br}
\affiliation{Departamento de Engenharia de Teleinform\'atica, Universidade Federal do Cear\'a, Fortaleza, Cear\'a, 60440, Brazil}
\affiliation{Applied and Computational Mathematics Division, National Institute of Standards and Technology, Boulder, Colorado, 80305, USA}


\date{\today}

\begin{abstract}
  Quantum state tomography aims to determine the quantum state of a
  system from measured data and is an essential tool for quantum
  information science.  When dealing with continuous variable quantum
  states of light, tomography is often done by measuring the field
  amplitudes at different optical phases using homodyne detection.
  The quadrature-phase homodyne measurement outputs a continuous
  variable, so to reduce the computational cost of tomography,
  researchers often discretize the measurements.  
  We show that this can be done without significantly
  degrading the fidelity between the estimated state and
    the true state.  This paper studies different
  strategies for determining the histogram bin widths. We show that
  computation time can be significantly reduced with little
  loss in the fidelity of the estimated state when
  the measurement operators corresponding to each histogram bin are
  integrated over the bin width.
\end{abstract}

\maketitle


\section{Introduction}
\label{intro}
Quantum information science and engineering have advanced to the point
where rudimentary quantum computers are available in the laboratory
and commercially~\cite{kandala2017, Linke2017, Monk2017, Denchev2016}.
However, further advancing quantum technologies requires improvements
in the fidelities of basic operations.  Consequently, more precise and
efficient reconstruction and diagnostic tools for estimation of quantum
states~\cite{Vogel1989, Smithey1993, Dunn1995, Banaszek1999,
  Banaszek2000, White2002, Ourjoumtsev2007, Neergaard2006},
processes~\cite{Chuang1997, Poyatos1997, Altepeter2003, Dariano1998,
  Nielsen1998, Mitchell2003, Obrien2004,Kupchak2015}, and
measurements~\cite{Luis1999, Fiurasek2001, Dariano2004, Lundeen2009}
are essential. Quantum tomographic techniques for optical quantum
states of light have become standard tools because quantum light
sources are essential for implementations of continuous-variable (CV)
quantum computation and communication~\cite{Lloyd1999, Gottesman2001,
  Bartlett2002, Jeong2002, Ralph2003}.  These source are also
extensively exploited in quantum cryptography~\cite{Ralph1999,
  Hillery2000, Silberhorn2002, Pirandola2008, Luiz2017}, quantum
metrology~\cite{Eberle2010, Demkowicz2013}, state
teleportation~\cite{Vaidman1994, Braunstein1998, He2015}, dense
coding~\cite{Braunstein2000, Lee2014} and cloning~\cite{Cerf2000,
  Braunstein2001}.

In the quantum state tomography studied here, one performs a
measurement on each member of a collection of quantum systems,
prepared in the same unknown state. Each system is measured in a basis
chosen from a complete set of measurements. The goal is to estimate
the unknown state from the measurements results.  This estimation can
be done by different methods, but we study Maximum Likelihood
Estimation (MLE), which finds among all possible states, the one that
maximizes the likelihood function.  The likelihood function
  computes for any state the probability, according to that state, of
obtaining the observed data.

Quantum homodyne tomography is one of the most popular optical
tomography techniques available~\cite{Lvovsky2004}. It rapidly became
a versatile tool and has been applied in many different quantum optics
experimental settings since it was proposed by Vogel and Risken in
1989~\cite{Vogel1989} and first implemented by Smithey \textit{et al.}
in 1993~\cite{Smithey1993}. This technique permits one to characterize
an optical quantum state by analyzing multiple phase-sensitive
measurements of the field quadratures.

A homodyne measurement generates a continuous value.  It is a
  popular practice to discretize the measurement result, because this
  can considerably reduce the size of the data and expedite the
  reconstruction calculation.  However the discretization necessarily
  loses information contained in the original measurements.
 How should we choose a discretization strategy such that the bins are not too small nor too large? Larger bins will reduce calculation time and memory, but
smaller bins will provide a better representation of the underlying distribution.
 
In this paper, we use numerical experiments to simulate optical
homodyne tomography and perform maximum likelihood tomography on the
data with and without discretization. When choosing a quadrature bin
width, we use and compare two different strategies: Scott's
rule~\cite{Scott2010} and Leonhardt's formula~\cite{Leonhardt1996}.
The paper is divided as follow: we begin by reviewing maximum
likelihood in homodyne tomography in Section \ref{MLE}. Then, in
Section \ref{numerical-experiments}, we describe our numerical
experiments. Next, we discuss the estimation of the mean photon number
from the quadrature measurements in Section
\ref{sec-photon-estimation}. In Section \ref{results} we present our
results, and we make our concluding remarks in Section
\ref{conclusion}.

\section{Maximum likelihood in homodyne tomography}
\label{MLE}
Let us consider $N$ quantum systems, each prepared in an optical state
described by a density matrix $\rhotrue$. In each experimental trial
$i$, we measure the field quadrature of one of the systems at some
phase $\theta_i$ of a local oscillator, i.e.\ a reference system
prepared in a high amplitude coherent state.  Each measurement is
associated with an observable
$\hat{X}_{\theta_i} = \hat{X} \cos \theta_i + \hat{P} \sin \theta_i$,
where $\hat{X}$ and $\hat{P}$ are analogous to mechanical position and
momentum operators, respectively. For a given phase $\theta_i$, we
measure a quadrature value $x_i$, resulting in the data
$\{(\theta_i, x_i)| i = 1, \ldots, N\}$.

The outcome of the $i$-th measurement is associated with a
positive-operator-valued measure (POVM) element
$\Pi (x_i|\theta_i) = \Pi_i$. Given the data, the likelihood
of a candidate density matrix $\rho$ is
\begin{eqnarray}
  \mathcal{L} (\rho)= \prod_{i=1}^{N} \mathrm{Tr} (\Pi_i \rho),
  \label{eq-likelihood}
\end{eqnarray}
where $\mathrm{Tr}(\rho \Pi_i)$ is the probability density, when
measuring with phase $\theta_i$, to obtain outcome $x_i$, according to
the candidate density matrix $\rho$.

MLE searches for the density matrix that maximizes the likelihood
in Eq.~(\ref{eq-likelihood}). It usually is
more convenient to maximize the logarithm of the likelihood (the
``log-likelihood''):
\begin{eqnarray}
  L (\rho) = \ln \mathcal{L} (\rho)= \sum_{i=1}^{N} \ln [\mathrm{Tr} (\Pi_i \rho)],
\end{eqnarray} 
which is maximized by the same density matrix as the likelihood. The
MLE is essentially a function optimization problem, and since the
log-likelihood function is concave, approximate convergence to a
unique solution will be achieved by most iterative optimization
methods.

In our numerical simulations, we use an algorithm for likelihood
maximization that begins with iterations of the $R\rho R$
algorithm~\cite{Rehacek2007} followed by iterations of a regularized
gradient ascent (RGA) algorithm. We switch from one algorithm to another 
because a slow-down is observed in the $R\rho R$ algorithm
after about $(t+1)^2/4$ iterations, where $t+1$ is the Hilbert
  space dimension. In the RGA, $\rho^{(k+1)}$ is parametrized as
\begin{equation}
  \rho^{(k+1)}=\frac{\left(\sqrt{\rho^{(k)}}+A\right)\left(\sqrt{\rho^{(k)}}+A^{\dagger}\right)}{\mathrm{Tr}\left[\left(\sqrt{\rho^{(k)}}+A\right)\left(\sqrt{\rho^{(k)}}+A^{\dagger}\right)\right]},
  \label{eq-rho-k+1}
\end{equation}
where $\rho^{(k)}$ is the density matrix found by the previous
iteration, and $A$ may be any complex matrix of the same dimensions as
$\rho$. Eq.~(\ref{eq-rho-k+1}) ensures that $\rho^{(k+1)}$ is a
density matrix for any $A$. We then use sequential quadratic
  programming optimization strategy \cite{Nocedal2006} in which $A$
is chosen to maximize the quadratic approximation of the
log-likelihood subject to $\text{Tr}(AA^{\dagger})\leq u$, where $u$
is a positive number adjusted by the algorithm to guarantee that the
log-likelihood increases with each iteration. To halt the iterations,
we use the stopping criterion of \cite{Glancy2012},
$L(\rho_{\text{ML}})-L(\rho^{(k)})\leq 0.2$, where
$L(\rho_{\text{ML}})$ is the maximum of the log-likelihood, which
  ensures convergence to a state whose log-likelihood is very close to
  the maximum likelihood.

\section{Methods for Numerical experiments}
\label{numerical-experiments}
Our numerical experiments simulate single mode optical homodyne
measurements of three types of states: (1) superpositions of coherent
states of opposite phase $|-\alpha\rangle + |\alpha\rangle$ (called
``cat states''), (2) squeezed vacuum states and (3) Fock states.  Each
state is represented by a density matrix $\rho_{\mathrm{true}}$
represented in the photon number basis, truncated at $t$ photons. To
better simulate realistic experiments, these pure states are
  subject to a 0.05 photon loss by passing
through a medium with transmissivity of 0.95 before measurement.

In order to calculate the probability to obtain homodyne measurement
outcome $x$, when measuring state $\rho_{\mathrm{true}}$ with phase
$\theta$, we represent all states and measurements
  in the photon number basis of a Hilbert space truncated at $t$
  photons. If $|x\rangle$ is the 
x-quadrature eigenstate with eigenvalue $x$, and $U(\theta)$ is the
phase evolution unitary operator, then for an ideal homodyne
measurement, we have $\Pi(x|\theta) = U(\theta)^{\dagger} |x\rangle
\langle x| U(\theta)$. To include photon detector inefficiency, we
replace the projector with $\Pi(x|\theta) = \sum_{i=1}^{n}
E_i(\eta)^{\dagger} U(\theta)^{\dagger} |x\rangle \langle x| U(\theta)
E_i(\eta)$, where $\eta$ is the detection efficiency and $E_i(\eta)$
are the associated Kraus operators \cite{Lvovsky2004}.  Typical
state-of-the-art homodyne detection systems have efficiency $\eta \sim
0.9$, so we use this value in our simulations. Using this strategy, we
are able to correct for the detector inefficiency as we estimate the
state. We use rejection sampling from the distribution given by
$P(x|\theta)$ to produce random samples of
homodyne measurement results~\cite{Kennedy1980}.

To choose the phases at which the homodyne measurements are performed,
we divide the upper-half-circle evenly among $m$ phases between 0 and
$\pi$ and measure $N/m$ times at each phase, where $N$ is the total
number of measurements. In all simulations, we use $m=20$ and
$N = 20,000$. To secure a single maximum of the likelihood function,
we need an informationally complete set of measurement operators,
which can be obtained if we use $t+1$ different phases to reconstruct
a state that contains at most $t$ photons~\cite{Leonhardt1997}.

To quantify the similarity of the reconstructed state $\rho$ to the
true state $\rhotrue$ we use the fidelity
\begin{eqnarray}
  F = \mathrm{Tr} \sqrt{\rho^{1/2}\, \rhotrue \, \rho^{1/2}}.
\end{eqnarray}
We report fidelities for four different
situations: (i) the state is reconstructed using the continuous values
of homodyne measurement results, that is without discretization; (ii)
the state is reconstructed with chosen bin widths (iii) the state is
reconstructed with bin widths given by Scott's rule~\cite{Scott2010};
and (iv) the state is reconstructed with bin widths suggested by
Leonhardt in~\cite{Leonhardt1997}. We only consider histograms with
contiguous bins of equal width.

In 1979 Scott derived a formula recommending a bin width for
discretizing data sampled from a probability density function $f$ for
a single random variable $x$. The recommended bin width is
\begin{eqnarray}
  h^{\star} = \left[ \frac{6}{s \int_{-\infty}^{\infty} f'(x)^2 dx} \right]^{1/3},
  \label{eq-hstar}
\end{eqnarray}
where the first and second derivatives of $f$ must be continuous and
bounded and $s$ is the sample size. Because one does not know $f$ in
an experiment we assume a normal distribution. For a normal $f$ we
have
\begin{eqnarray}
  \int_{-\infty}^{\infty} f'(x)^2 dx = \frac{1}{4 \sqrt{\pi} \sigma ^3},
  \label{eq-intnormaldist}
\end{eqnarray}
where $\sigma$ is the distribution's standard deviation.  Combining
Eqs.~(\ref{eq-hstar}) and (\ref{eq-intnormaldist}), we obtain the
recommended bin width for a normal distribution:
\begin{eqnarray}
  h = 3.5 \, \sigma \, s^{-1/3}.
  \label{eq-scott}
\end{eqnarray}
This formula is known as Scott's rule, and is optimal for 
  estimating $f$ (minimizing total mean-squared error) at each phase
if the data is normally distributed. In our simulations we compute a
bin size separately for each phase's quadrature measurements, and we
use the unbiased sample standard deviation in place of $\sigma$.

Although Scott's rule is optimal for each phase, it may not be optimal
for homodyne tomography because we are estimating the density matrix
rather than each phase's quadrature distribution individually.  Also
many interesting optical states do not have normal quadrature
distributions, for example our cat states.  In fact, one might expect
that the bin width should be related to the number of photons in a
quantum state because higher photon number states have more narrow
features in their quadrature distributions, which should not be
washed-out by the discretization.

Leonhardt states that if we desire to reconstruct a density matrix of
a state with $n$ photons, we need a bin width narrower than $q_n/2$,
where $q_n$ is given by
\begin{eqnarray}
  q_n = \frac{\pi}{\sqrt{2 n + 1}}.
  \label{eq-leonhardt}
\end{eqnarray}
This result was obtained from a semiclassical
approximation for the amplitude pattern functions in quantum state
sampling~\cite{Leonhardt1996}. Leonhardt recommends using the maximum
photon number in Eq.~(\ref{eq-leonhardt}), however many states have no
maximum photon number, and whether a state has a maximum
photon number is not possible to learn with certainty from tomography.
Instead, we have tested using the photon number $t$ at which the
reconstruction Hilbert space is truncated and an estimate of
the mean photon number $\overline{\langle \hat{n} \rangle}$ in
Eq.~(\ref{eq-leonhardt}).  The truncation $t$ must be chosen in
advance to be large enough that the probability that $\rhotrue$
contains more than $t$ photons is very small.  We estimate the mean 
photon number from the quadrature measurements as described in the next section.

\section{Estimating mean photon number}
\label{sec-photon-estimation}
In order to use Leonhardt's advice for choosing the histogram bin
width, we need to estimate the mean number $\langle \hat{n} \rangle$
of photons in the measured state from the
phase-quadrature data set. We use the estimator given in
Refs.~\cite{Hradil2,Munroe1996}.  To obtain this estimator, we first
compute the mean value of $(\hat{X}_{\theta})^{2}$, averaged over
$\theta$, treating $\theta$ as if it is random and uniformly
distributed between $0$ and $\pi$.
\begin{equation}
  \langle (\hat{X}_{\theta})^{2} \rangle = \langle \hat{X}^{2}\cos^{2}\theta + (\hat{X}\hat{P}+\hat{P}\hat{X})\cos\theta\sin\theta + \hat{P}^{2}\sin^{2}\theta \rangle
\end{equation}
The phase $\theta$ is independent of $\hat{X}$ and $\hat{P}$, so we
can compute the expectation over $\theta$ as
\begin{align}
  \langle (\hat{X}_{\theta})^{2} \rangle &= \Big\langle \int_{0}^{\pi} (\hat{X}^{2}\cos^{2}\theta + (\hat{X}\hat{P}+\hat{P}\hat{X})\cos\theta\sin\theta \nonumber \\
                                         & \qquad \qquad + \hat{P}^{2}\sin^{2}\theta) \mathrm{Prob}(\theta) \mathrm{d}\theta \Big\rangle \\
  \langle (\hat{X}_{\theta})^{2} \rangle &= \Big\langle \int_{0}^{\pi} (\hat{X}^{2}\cos^{2}\theta + (\hat{X}\hat{P}+\hat{P}\hat{X})\cos\theta\sin\theta \nonumber \\
                                         & \qquad \qquad + \hat{P}^{2}\sin^{2}\theta) \frac{1}{\pi} \mathrm{d}\theta \Big\rangle \\
                                         &= \frac{1}{2}\left\langle \hat{X}^{2} + \hat{P}^{2} \right\rangle.
\end{align}
Because the photon number operator is
\begin{equation}
  \hat{n} = \frac{1}{2}\left(\hat{X}^{2}+\hat{P}^{2}-1\right),
\end{equation}
we obtain
\begin{equation}
  \langle\hat{n}\rangle = \langle \hat{X}_{\theta}^{2}\rangle-\frac{1}{2}. 
\end{equation}
We estimate $\langle \hat{n} \rangle$ by computing the sample
mean of all quadrature measurements~\cite{Hradil2,Munroe1996}:
\begin{equation}
  \overline{\langle \hat{n} \rangle} = \frac{1}{N} \sum_{i=1}^{N}x_{i}^{2} - \frac{1}{2},
  \label{eq-photon-estimation}
\end{equation}
where the bar above $\overline{\langle\hat{n}\rangle}$
distinguishes the true mean photon number from our estimate of the
mean photon number.  Note that when $\theta$ is uniformly
distributed over $[0,\pi)$, the individual values of $\theta$ are not
needed to compute $\overline{\langle \hat{n} \rangle}$.

\section{Results}
\label{results}
\begin{figure*}
  \includegraphics[width=0.78\textwidth]{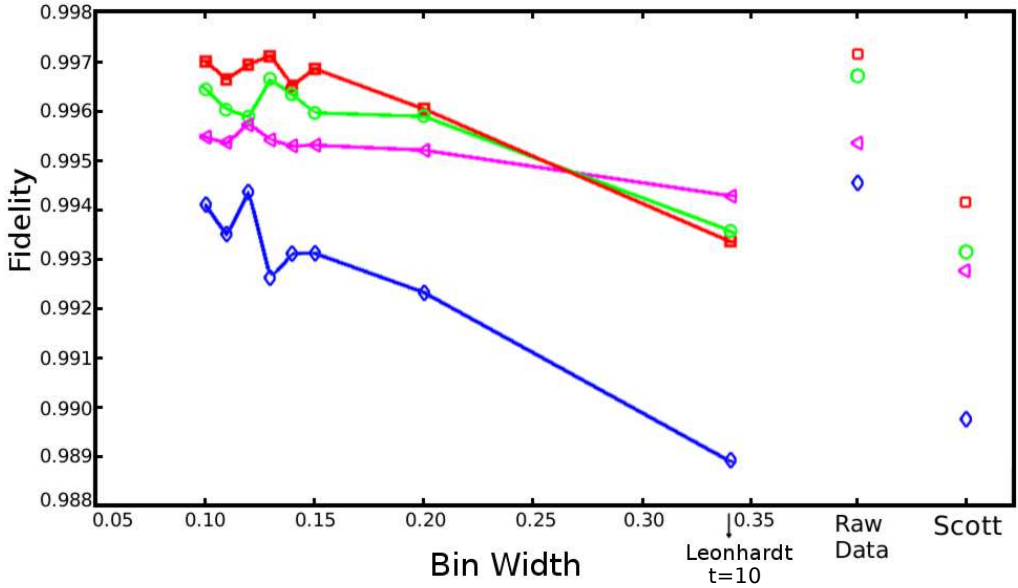}
  \caption{Fidelities between estimated states and true states as
    functions of the bin width for a cat state with amplitude
    $\alpha=1$ and photon loss of 0.05. The Hilbert space is truncated
    at $t=10$ photons. Each set of points with the same color and
    marker shape corresponds to a different data set. The mean bin
    width for Scott method is $0.35$.}
  \label{fig-methods_fidelity_singledata}
\end{figure*}

\begin{figure*}
  \includegraphics[width=0.78\textwidth]{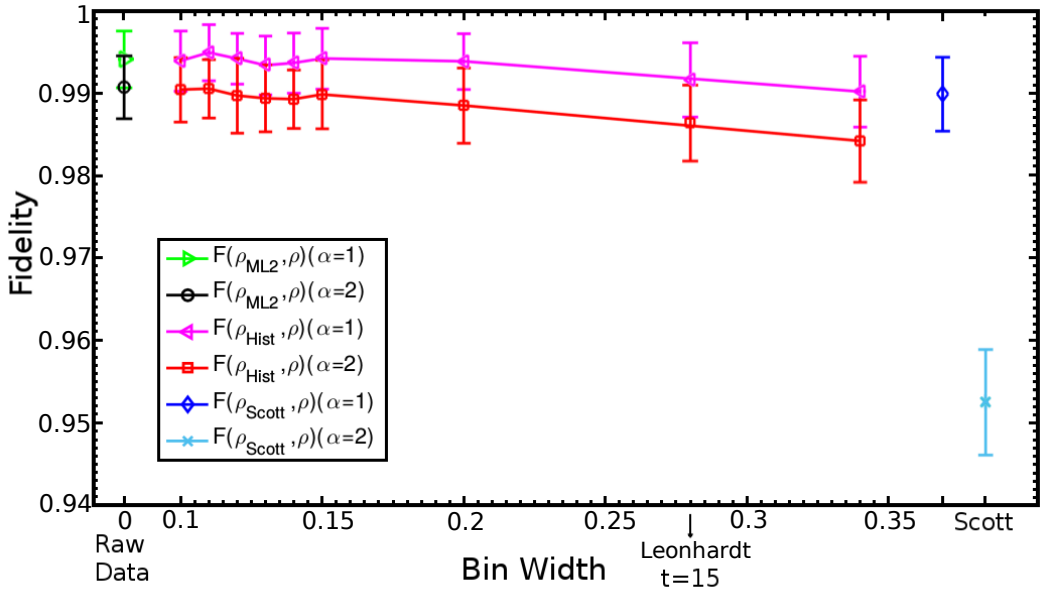}
  \caption{Average fidelity as a function of the bin width for cat
    states with amplitudes $\alpha=1$ and $\alpha=2$. The Hilbert
    space is truncated at $t=15$ photons. The mean bin widths for
    Scott's method are $0.35$ ($\alpha=1$ cat state) and $0.64$
    ($\alpha=2$ cat state).}
  \label{fig-fidelity_vs_binwidth_15_photons_catstate}
\end{figure*}

\begin{figure*}
  \includegraphics[width=0.78\textwidth]{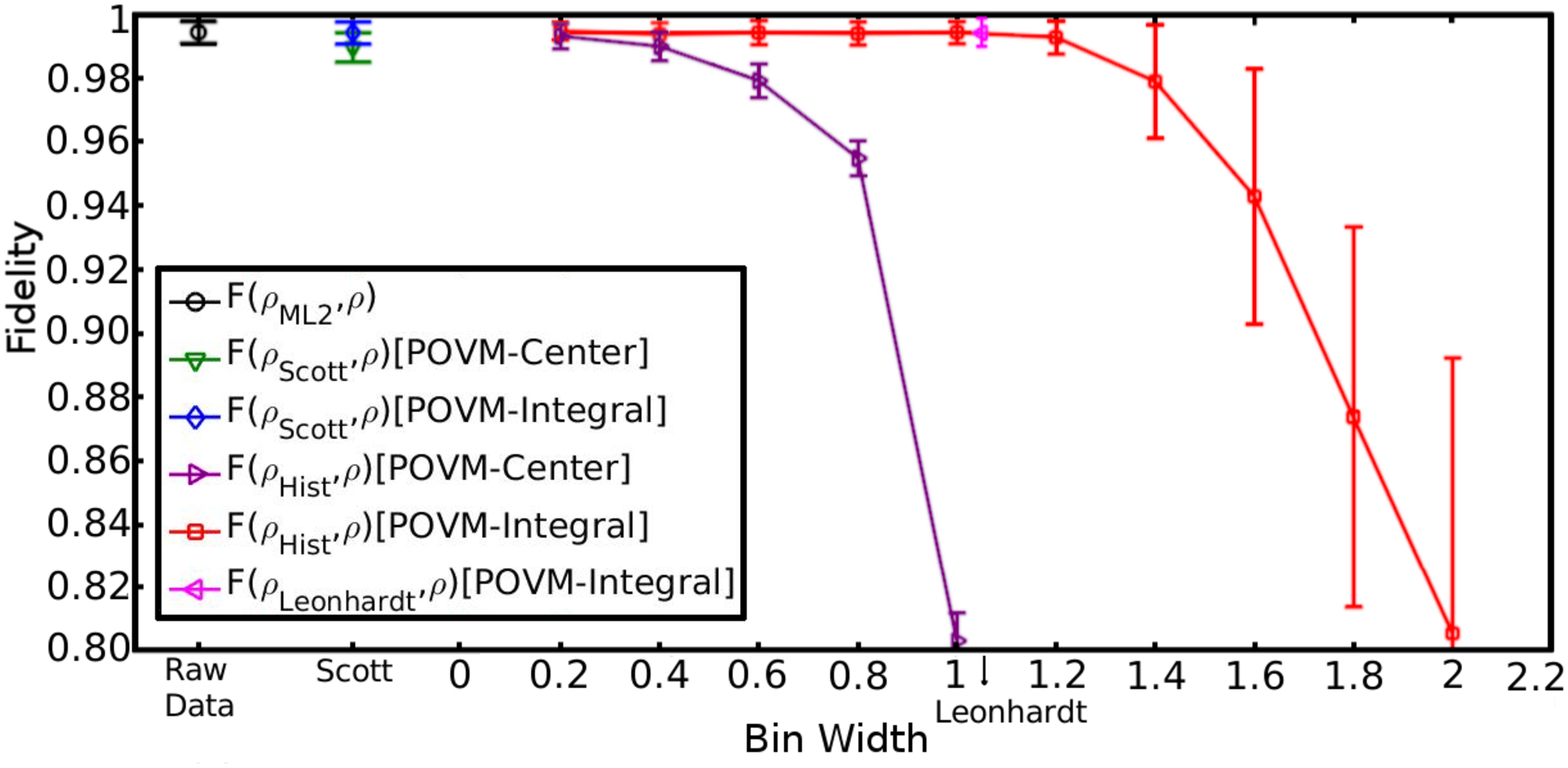}
  \caption{Average fidelity as a function of the bin width for a cat
    state with amplitude $\alpha = 1$. The Hilbert space is truncated
    at $t=10$ photons. For this state, $\langle n \rangle = 0.6093$, and 
    $\overline{\langle \hat{n} \rangle}=0.6109$, giving a bin width by
    Leonhardt's formula of $1.05$. The mean bin width for Scott's method 
    is $0.35$.}
  \label{fig-Fidelity_vs_binwidth_catstate_Mph_10_alpha_1}
\end{figure*}

To study the performance of various discretization strategies, we
compute fidelities between the true state and the states estimated
with the different strategies. Below $\rho_{\mathrm{ML2}}$ represents
the state estimated without discretization, $\rho_{\mathrm{Hist}}$ is
estimated with histogram bins of specified width chosen arbitrarily,
$\rho_{\mathrm{Scott}}$ is estimated with bin widths chosen according
to Scott's rule, and $\rho_{\mathrm{Leonhardt}}$ is estimated with
Leonhardt's bin widths. To make the graphs below, for each choice of
parameters, we simulate 100 tomography experiments, making 100 density
matrix estimates.  The graphs show the arithmetic mean of the 100
fidelities of the reconstructed states. The half width of the
error bars are the sample standard deviations of
the 100 fidelities.

Our first results are shown in
Fig.~\ref{fig-methods_fidelity_singledata}. The state considered is a
cat state with $\alpha = 1$, where $\alpha$ is the amplitude of the
coherent state in the superposition.  The state is reconstructed in a
Hilbert space truncated at $t=10$ photons. (The probability that the
$\alpha=1$ state has more than 10 photons is $3.8 \times 10^{-10}$.)
Scott's method finds a different optimal bin width for each phase
considered, so we report the mean bin width averaged over the 20
phases in these cases.  Here the mean bin width for Scott's method is
$0.35$.  When choosing a bin width, we use values up to
$0.34$, the width we obtain when we use
Eq.~(\ref{eq-leonhardt}) for $t=10$, the number of photons at which we
truncated the Hilbert space.  In all cases, each bin's measurement
operator represents the measurement as if it occurred at the center of
each bin. In Fig~\ref{fig-methods_fidelity_singledata}, each set of
points corresponds to a different data set. We see that different data 
sets had similar behavior as we changed the bin
size. As we can see in this figure, the highest fidelities occur when
we do not use discretization, as expected.  We also see that smaller
bin widths result in higher fidelities. However, even the largest bin
widths tested result in a fidelity loss of only 0.005 compared to the
raw data.

The next set of results is presented in
Fig.~\ref{fig-fidelity_vs_binwidth_15_photons_catstate}, where we show
average fidelities as a function of the bin width for cat states with
amplitudes $\alpha=1$ and $\alpha=2$. The states are reconstructed in
a $t=15$ photons Hilbert space. The $\alpha=2$ state has probability
of $3.3 \times 10^{-7}$ to contain more than 15 photons.  The fidelity
for an $\alpha=1$ cat state is always greater than the fidelity for a
$\alpha=2$ cat state, including the case when we do not use
discretization. This is expected, because a $\alpha = 2$ state
requires more parameters to effectively describe its density matrix in
the photon number basis, so for a given amount of data, there is
greater statistical uncertainty.

For a given bin width the fidelity of the
  $\alpha=2$ cat state estimates is always lower than the fidelity for
  the $\alpha=1$ cat state estimates. This is also expected because
the $\alpha = 2$ state has more wiggles in its probability
distribution, so more information is lost when the bins are
larger. The average bin width used by Scott's method is $0.35$ for the
$\alpha=1$ cat state, and $0.64$ for the $\alpha=2$ cat state, which
results in significant fidelity loss. Leonhardt's width indicated in
Fig.~\ref{fig-fidelity_vs_binwidth_15_photons_catstate} is obtained by
using $t=15$ in place of $n$ in Eq.~(\ref{eq-leonhardt}).

\begin{figure*}
  \includegraphics[width=0.78\textwidth]{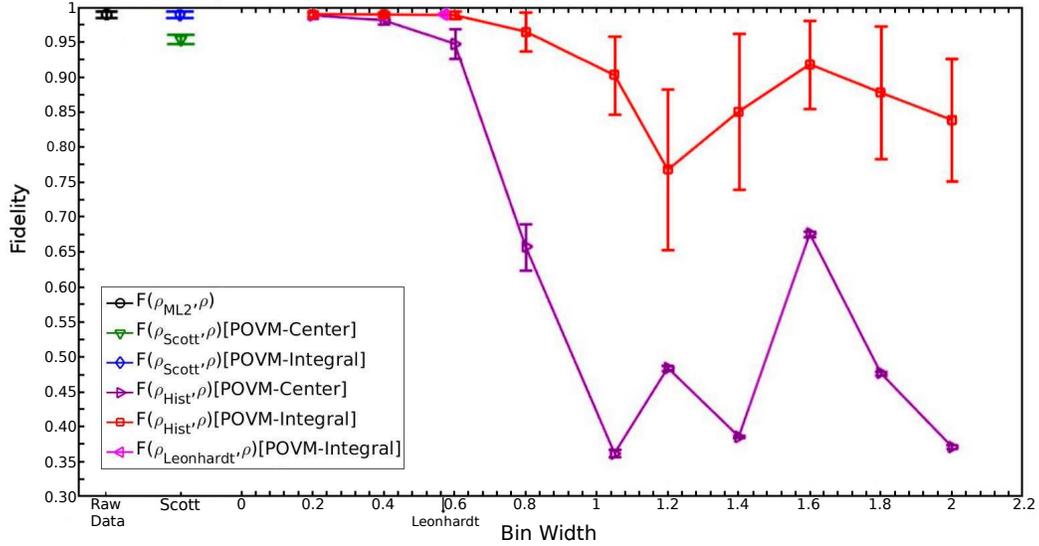}
  \caption{Average fidelity as a function of the bin width for a cat
    state with amplitude $\alpha = 2$. The Hilbert space is truncated
    at $t=15$ photons. For this state, $\langle n \rangle = 3.1978$, and 
    $\overline{\langle \hat{n} \rangle}=3.1983$, giving a bin width by
    Leonhardt's formula of $0.58$.  The mean bin width for Scott's
    method is $0.64$.}
  \label{fig-Fid_vs_binwidth_catstate_alpha_2_Mph_15}
\end{figure*}

\begin{figure*}
  \includegraphics[width=0.78\textwidth]{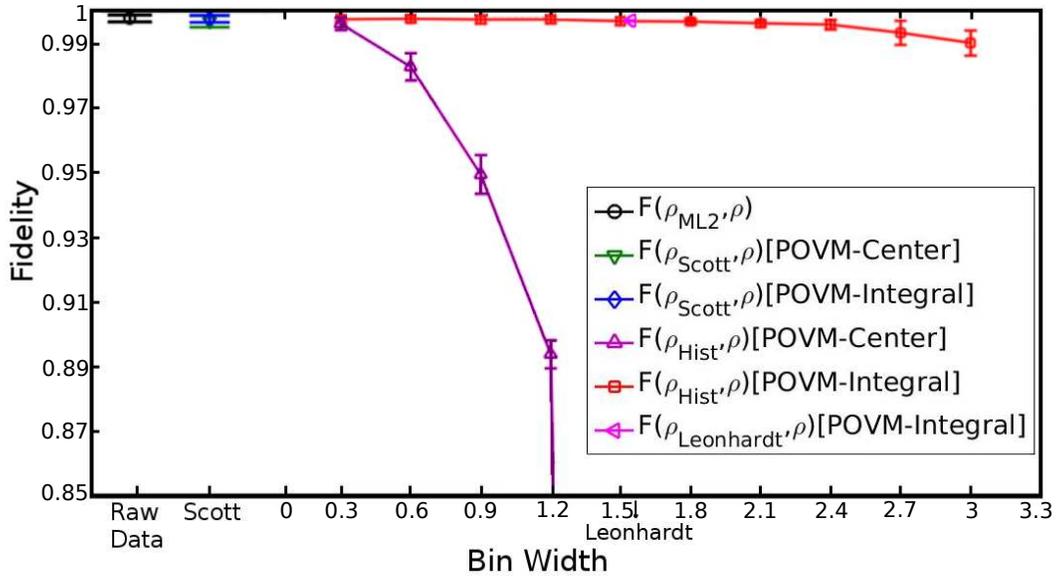}
  \caption{Average fidelity as a function of the bin width for a
    squeezed vacuum state whose squeezed quadrature has a variance 3/4
    of the vacuum variance. The Hilbert space is truncated at $t=10$
    photons. For this state, $\langle n \rangle = 0.0167$, and
    $\overline{\langle \hat{n} \rangle}=0.0162$, giving a bin width by
    Leonhardt's formula of $1.54$.  The mean bin width for Scott's
    method is $0.25$.}
  \label{fig-squeezed_vacuum_variance_075_Mph_10}
\end{figure*}

Until now, as mentioned before, every measurement outcome in a given
bin has been associated with the measurement operator for the
quadrature value at the center of that bin. That is, the
  measurement operator $\Pi_{i}$ associated with bin $i$ would give the
  probability density of obtaining a measurement result at the center
  of bin $i$ when computing $\mathrm{Tr}(\Pi_{i}\rho)$. Although this
may be a useful approximation for very small bins, to improve our
analysis, we now change each bin's measurement operator so that it
represents a measurement that occurs anywhere in the bin. To obtain
these new operators, we numerically integrate the measurement
operators over the width of each histogram bin. With these
  integrated measurement operators, computing
  $\mathrm{Tr}(\Pi_{i}\rho)$ will gives the probability to obtain a
  measurement result anywhere in bin $i$. We identify each case by
adding $[$POVM-center$]$ and $[$POVM-integral$]$ to the legends in the
graphs.

We also add to our analysis the use of the mean photon number estimate
in Leonhardt's formula, and we calculate the fidelity between
$\rhotrue$ and the state $\rho_{\mathrm{Leonhardt}}$ estimated using
the resulting bin width. Recall that Leonhardt recommends that the bin
width should be smaller than the one calculated using the maximum photon 
number in Eq.~(\ref{eq-leonhardt}) but here we use the estimate of 
the mean photon number instead.

Figs.~\ref{fig-Fidelity_vs_binwidth_catstate_Mph_10_alpha_1}
and~\ref{fig-Fid_vs_binwidth_catstate_alpha_2_Mph_15} show average
fidelities as functions of the bin width for cat states with
amplitudes $\alpha=1$ and $\alpha=2$,
respectively. Fig.~\ref{fig-squeezed_vacuum_variance_075_Mph_10}
  examines a squeezed vacuum state whose squeezed quadrature has a
  variance 3/4 of the vacuum variance.  Note for the cat states, as
  $\alpha$ increases, Scott's bin width also increases, which is
  certainly undesirable because the quadrature distributions contain
  more fine structure.  These graphs show that integrating the
  measurement operators over the width of each bin considerably
  improves the fidelities for all cases. We can also see that
  Leonhardt's suggestion using the estimated mean photon number can be
  safely used as the upper bound for the bin width.

\begin{figure*}
  \includegraphics[width=0.78\textwidth]{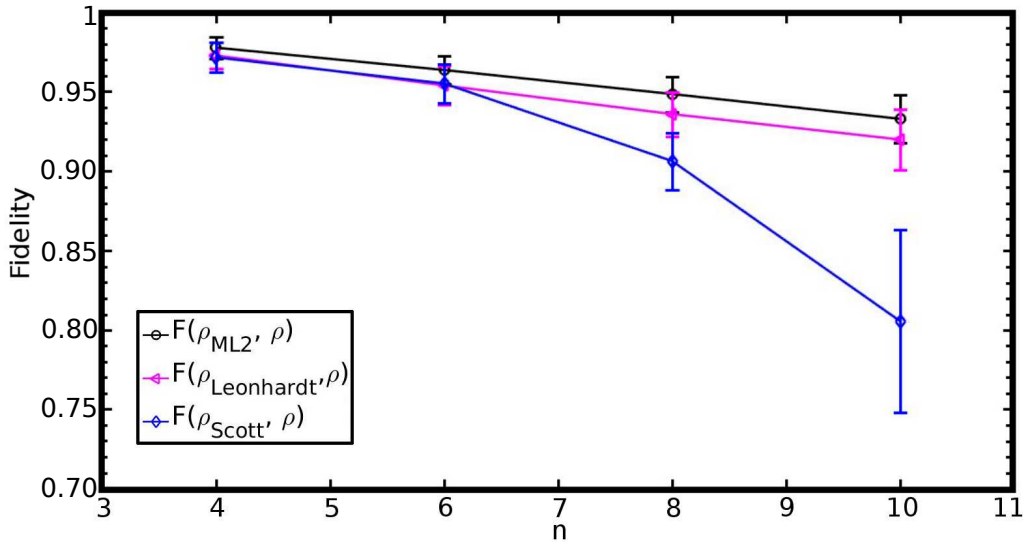}
  \caption{Average fidelities as functions of bin width for Fock
    states with different numbers of photons $n$. The bin widths from
    Leonhardt's formula are 0.56 ($n=4$), 0.46 ($n=6$), 0.40 ($n=8$),
    and 0.36 ($n=10$). The mean bin width from Scott's method are 0.69
    ($n=4$), 0.82 ($n=6$), 0.95 ($n=8$), and 1.05 ($n=10$).}
  \label{fig-fock}
\end{figure*}

As seen in Eq.~(\ref{eq-scott}), Scott's rule gives bin widths
proportional to the sample standard deviation. Since states with a
higher number of photons can have higher standard deviations,
Scott's method will produce larger bin widths. This is
undesirable because pure states containing many photons have very
fine features in their quadrature distributions. On the other
hand, we expect Leonhardt's method to perform better
because it uses the estimated mean number of photons
to calculate the bin width. We can clearly see the expected behavior
of both methods for higher numbers of photons in Fig.~\ref{fig-fock},
where we have used Fock states to check our claim.

\begin{figure*}
  \includegraphics[width=0.7\textwidth]{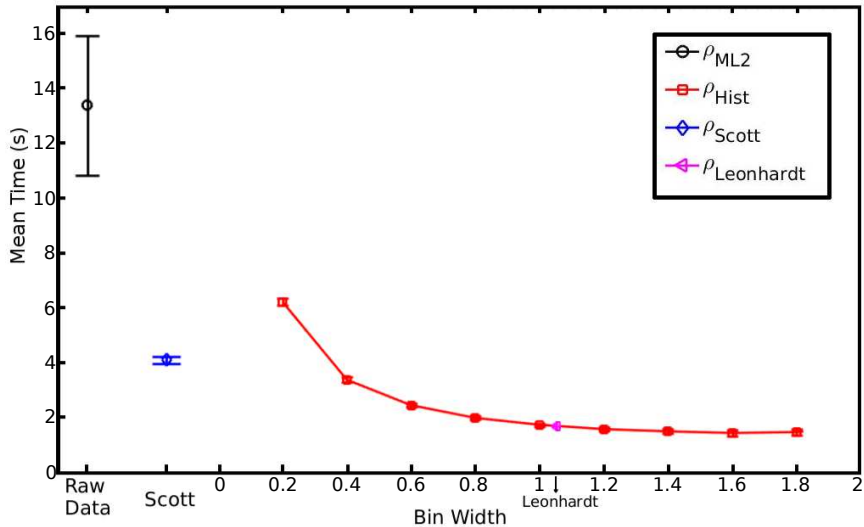}
  \caption{Average reconstruction time as a function
    of the bin width for a cat state with amplitude $\alpha = 1$. The
    Hilbert space is truncated at $t=10$ photons. The mean bin width
    for Scott's method is $0.35$, and the bin width given by
    Leonhardt's formula is $1.05$.}
  \label{fig-time-catstate}
\end{figure*}

\begin{figure*}
  \includegraphics[width=0.7\textwidth]{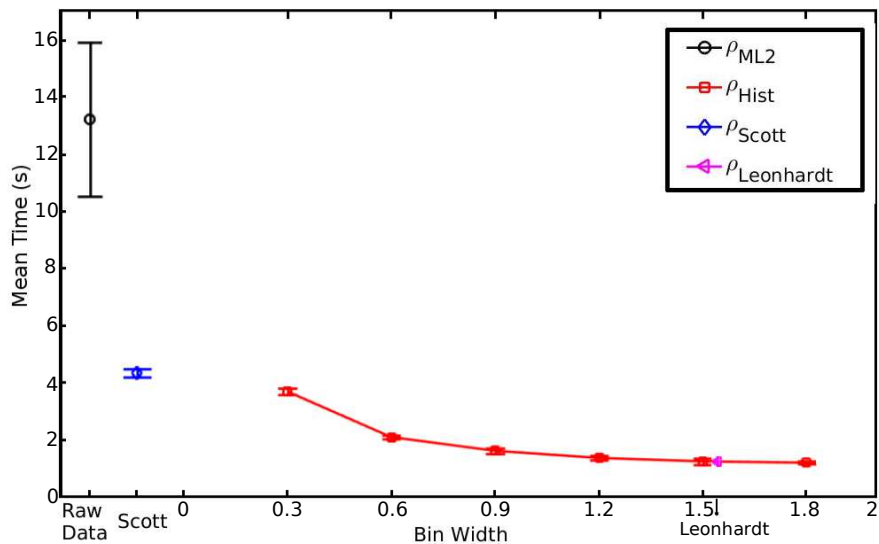}
  \caption{Average reconstruction time as a function
    of the bin width for a squeezed vacuum state whose squeezed
    quadrature has a variance 3/4 of the vacuum variance. The Hilbert
    space is truncated at $t=10$ photons. The mean bin width for
    Scott's method is $0.25$, and the bin width given by Leonhardt's
    formula is $1.54$.}
  \label{fig-time-squeezed}
\end{figure*}

All of the discretization methods considered here give much faster
fidelity estimates, as we can see in Figs.~\ref{fig-time-catstate}
and~\ref{fig-time-squeezed}, with no significant loss of fidelity
between the estimated states and the true states. The
average times reported here include any calculations required to
determine the desired bin width from the original homodyne data, the
construction of histograms, and the ML density matrix estimation. All
the simulations were carried out in a dual-core computer running at
3.7 GHz with 4 GB of RAM.

\section{Conclusion}
\label{conclusion}

We have used idealized numerical experiments to generate simulated
data, performed maximum likelihood tomography on data sampled
from cat states and squeezed vacuum states with and without
discretization, and estimated the fidelities between
the reconstructed states and the true state. We used two different
methods to choose the bin width: Scott's and Leonhardt's methods. We
studied using measurement operators calculated using the
quadrature exactly at the center of each bin and integrating the
measurement operators along the length of the bin. 

Scott's method calculates an optimal bin width, for each phase, based
on the size and the standard deviation of the sample. This method
works well for Gaussian states and states with small numbers of
photons.  States with higher number of photons have quadrature
  distributions with higher standard deviations, giving bigger bin
widths for each phase. We implemented Scott's method for Gaussian
distributions, but if one has prior knowledge about the state and its
distribution, one could tailor Scott's rule by using more appropriate
distributions in Eq.~(\ref{eq-hstar}).

Leonhardt's method recommends a bin width narrower than $q_n/2$, where
$q_n$ decreases with the square root of the number of photons in the
state being reconstructed.  Since, in a real experiment, we do not
know the mean number of photons in the state considered, we estimate
the mean photon number from the quadrature measurement results. We
have found that the method to find the mean number of photons
from the quadrature measurement results gives accurate results. We
checked that by comparing the estimated mean number of photons with
the true mean number of photons for the cat states and squeezed vacuum
states. We also have found that integrating the measurement operators
over the width of each histogram bin significantly improves the
fidelity. Using this strategy, Leonhardt's formula safely establishes
an upper bound to the bin width, and both methods provides a faster
statistical estimation without losing too much information.

\begin{acknowledgments}
  We thank Karl Mayer and Omar Magana Loaiza for helpful comments on
  the manuscript.  H. M. Vasconcelos thanks the Schlumberger
  Foundation's Faculty for the Future program for financial
  support. J. L. E. Silva thanks Coordena\c c\~ao de Aperfei\c
  coamento de Pessoal de N\'ivel Superior (CAPES) for financial
  support. This work includes contributions of the National Institute
  of Standards and Technology, which are not subject to
  U.S. copyright.
\end{acknowledgments}


%

%


\end{document}